\documentstyle[11pt,aaspp4]{article}   

\def\hst{{\it HST}}
\def\etal{{\it et al.}}
\def\Msun{M_{\odot}}

\def\F218W{{\rm F218W}}

\begin{document}

%
%

\title{Peculiar Multimodality on the Horizontal Branch \nl of the
Globular Cluster NGC 2808\footnote{Based on observations with
the NASA/ESA {\it Hubble Space Telescope}, obtained at the Space
Telescope Science Institute, which is operated by AURA, Inc., under
NASA contract NAS 5-26555.}}

\author{Craig Sosin}

\affil{Astronomy Department, University of California, Berkeley, CA
94720--3411}

\author{Ben Dorman}

\affil{Laboratory for Astronomy and Solar Physics, Code 681, \nl
NASA Goddard Space Flight Center, Greenbelt, MD 20771 \\ and \\
Dept.\ of Astronomy, University of Virginia, P.\ O.\ Box 3818, Charlottesville,
Virginia 22903-0818}

\author{S.\ George Djorgovski}

\affil{Astronomy Department, MS 105--24, California Institute of
Technology, \nl Pasadena, CA 91125}

\author{Giampaolo Piotto}

\affil{Dipartimento di Astronomia, Universit\`a di Padova, Vicolo
dell'Osservatorio 5, \nl I--35122 Padova, Italy}

\author{R.\ Michael Rich}

\affil{Astronomy Department, Columbia University, 538 W.\ 120th St., \nl
Box 43 Pupin, New York, NY 10027}

\author{Ivan R.\ King}

\affil{Astronomy Department, University of California, Berkeley, CA
94720--3411}

\author{James Liebert}

\affil{Steward Observatory, University of Arizona, Tucson, AZ 85721}

\author{E.\ Sterl Phinney}

\affil{Theoretical Astrophysics, MS 130--33, California Institute of
Technology, \nl Pasadena, CA 91125}

\author{Alvio Renzini}

\affil{Dipartimento di Astronomia, Universit\`a di Bologna, Cp.\ 596,
I--40100 Bologna, Italy}

\clearpage

\begin{abstract}
We present distributions of colors of stars along the horizontal
branch of the globular cluster NGC 2808, from {\it Hubble Space
Telescope} WFPC2 imaging in $B$, $V$, and an ultraviolet filter
(F218W).  This cluster's HB is already known to be strongly bimodal,
with approximately equal-sized HB populations widely separated in the
color--magnitude diagram.  Our images reveal a long blue tail with two
gaps, for a total of four nearly distinct HB groups.  These gaps are
very narrow, corresponding to envelope-mass differences of only
$\sim0.01 \Msun$.  This remarkable multimodality may be a signature of
mass-loss processes, subtle composition variations, or dynamical
effects; we briefly summarize the possibilities.  The existence of
narrow gaps between distinct clumps on the HB presents a challenge for
models that attempt to explain HB bimodality or other peculiar HB
structures.
\end{abstract}


\clearpage

%
%

\section{Introduction}

Despite decades of work, the causes of the distribution of colors of
horizontal-branch stars in globular clusters remain a mystery.  In
general, more-metal-rich clusters have red HBs, while metal-poor HBs
are bluer.  But there is a large degree of variation in HB
morphologies, and a number of cluster HBs do not have the color
expected for their metal abundance.  The distribution of observed
colors requires that up to $\sim0.3\Msun$ have been lost by each star
(Castellani \& Renzini 1968, Iben \& Rood 1970).  The mechanism for
this mass loss is not understood, though, and our lack of
understanding of this phenomenon pervades all discussion of this
subject.

A particularly strange piece of this puzzle has been the few clusters
that have bimodal HB distributions, including NGC 1851 (Stetson 1981),
2808 (Harris 1974), and 6229 (Borissova \etal\ 1997) at moderate
metallicities.  Perhaps even more surprising are the newly discovered
bimodal HBs in the metal-rich clusters NGC 6388 and 6441 (Rich \etal\
1997).  A set of extremely hot HB stars is also found in the
metal-rich open cluster NGC 6791 (Kaluzny \& Udalski 1992, Liebert,
Saffer, \& Green 1994), whose HB is predominantly red.  As well as
these red--blue bimodalities, a number of clusters appear to have gaps
along the blue HB tail, with the bluest stars distinguished from the
slightly cooler stars by an obvious break in the HB sequence.

In this paper we focus mainly on the blue HB of NGC 2808.  We present
new observations of the central part of the cluster, taken with the
Wide Field/Planetary Camera 2 (WFPC2) aboard the {\it Hubble Space
Telescope (HST)}.  These images contain a larger sample of
post-main-sequence stars in the cluster than any taken previously.
They show that the blue part of the cluster's HB is considerably more
extended than previously known, and reveal two narrow gaps along the
blue HB.  The overall distribution of HB stars is thus {\it
multi}-modal, with at least four significant populations.

The distinct clumps along the blue HB tail, and the narrow gaps that
separate them, present a special challenge for models of bimodal and
other peculiar HB morphologies.

\section{Observations and Reduction}

NGC 2808 was observed by the WFPC2 on 2 December 1995.  Several
exposures were taken through the F555W and F439W filters, which
approximate Johnson $V$ and $B$; total exposure times were 57 s and
510 s, respectively.  A longer series of exposures (totaling 3300 s)
was taken through F218W, a UV filter that peaks near $\rm 2180 \AA$.
The images were processed by the standard \hst\ pipeline; we then used
DAOPHOT (Stetson 1987) to identify stars and measure magnitudes.

The photometry was calibrated to Johnson $V$ and $B$ (for F555W and
F439W), and to the STMAG system (for F218W), using zeropoints and
transformations given by Holtzman \etal\ (1995).  The effects of
contamination in the UV image were removed according to the
prescription of Whitmore, Heyer, \& Baggett (1996).  Our calibration
is $\sim0.1$ mag fainter in $V$, and $\sim0.1$ mag bluer in $B-V$,
than that of Ferraro \etal\ (1990); Ferraro (1996, private
communication) has confirmed that his recalibration of their data is
consistent with these differences.

The $(V, B-V)$ color--magnitude diagram (CMD) is shown in Figure 1,
which contains over 27,000 objects.  The previously known portions of
the HB are visible near $(V = 16.3, B-V = 0.8)$ and $(V=17, B-V=0.1)$.
What is new in Fig.\ 1 is the extension of the blue tail of the HB,
all the way down to $V \simeq 21$.  Moreover, even from this diagram
it is clear that the blue tail is not uniformly populated:\ there are
discontinuities in the density of points near $V = 18$ and $V = 20$.

The structure of the blue HB becomes more apparent in the $(B,
\F218W-B)$ diagram in Figure 2, which contains 437 objects.  The stars
farthest to the left in the lower panel are at the bottom of the blue
tail in Fig.\ 1.  The stars near $(B = 18, \F218W-B = 1)$ correspond
to the brightest blue stragglers in Fig.\ 1.  The red HB clump and the
RGB are faint in $\F218W$, and do not appear in Fig.\ 2.  The upper
panel of Fig.\ 2 shows a histogram of $\F218W-B$ colors of HB stars.
There are three peaks, at $\F218W - B \simeq -0.5$, $-1.7$, and
$-2.4,$ and two gaps, near $\F218W - B = -1.9$ and $-1.3$ (not
entirely empty).  When the red HB clump is included, the HB of NGC
2808 is now seen to be {\it four} groups of stars along a single
sequence.

We ran completeness experiments using the F218W and $B$ images.  The
counts are virtually 100\% complete for HB stars brighter than $B=20,$
and are $\sim90\%$ complete for the bluest clump, averaged over the
observed area.  The small amount of incompleteness will not
dramatically change the relative heights of the three peaks of the
histogram in Fig.\ 2, and certainly could not be the cause of the
gaps.

The four groups consist of $\sim350,$ 275, 70, and 60 stars,
respectively, going from red to blue.  The Lee index $(B-R) / (B+V+R)$
is thus $\sim0.1$, in contrast to previous studies that have found
more red stars than blue (Byun \& Lee 1993).

\section{Physical Parameters along the HB}

We fitted the zero-age HB (ZAHB) and evolutionary models of Dorman,
Rood, \& O'Connell (1993) to both observed CMDs.  Newly computed
models extend the ZAHB to envelope masses $M_{\rm env} = 10^{-4}
\Msun$.  Stars with such a small envelope are essentially on the He
main sequence, since H-burning is negligible throughout their HB
evolution.  Models computed with more modern physical inputs make no
difference for this paper (cf.\ Yi, Lee \& Demarque 1993, Hill \etal\
1996).  We used Kurucz (1992) synthetic stellar fluxes to generate
model colors, and assumed $A_{\rm F218W} / E(B-V) = 8.74,$ using the
extinction curve of Cardelli \etal\ (1989).  The extension of the ZAHB
redward of the red HB clump is an artifact of the Kurucz-model colors;
colors produced by Bell (1997, private communication) do match the
clump.

We could not find a combination of $(m-M)_0$ and $E(B-V)$ that fit
both CMDs with the nominal filter zeropoints, probably due to
uncertainties in the calibration and in the transformation from the
theoretical to the observational plane.  However, by shifting the
zeropoints (within the uncertainties given by Holtzman \etal), we find
excellent fits with parameters ranging from $[(m-M)_0,E(B-V)] =
[15.25, 0.16]$ to $[15.4, 0.09]$, depending on how we divide the
zeropoint shift between the optical color and the F218W magnitude.
The canonical values are $14.8$ and $0.22,$ respectively (Ferraro
\etal\ 1990, Peterson 1993), but the new UV colors and revised $V$
magnitudes rule out those values, even with the maximum possible
shifts.  Note that most of the difference between the old and new
values is along the reddening vector.

In Figure 3 we show the model fit to the data for one fit (15.25 and
0.16), with the \hst\ $B-V$ colors shifted 0.07 mag to the red.  The
HB evolutionary tracks shown are for models with masses of 0.540,
0.530, 0.510, 0.495, and $0.4848\Msun$, going from red to blue, with
$M_{\rm core} = 0.4847 \Msun$.  The bluest track is for a star with
virtually no envelope, and thus represents a theoretical limit to the
HB.  The isochrone shown is for a population with [Fe/H] $= -1.48$ and
an age of 14 Gyr ({\it not} fitted to the data near the turnoff).  The
zero-age main sequence shown extends up to $1.5 \Msun$.

The two blue HB gaps are at $T_{\rm eff}$ near $17,000$ and $25,000$
K, and masses near 0.54 and $0.495\Msun$, respectively.  In both cases
the gap widths correspond to mass-loss differences of only $\sim0.01
\Msun$.  (The other gap, between the red and blue HBs, is wider:\
$\sim0.05\Msun$.)  The two intermediate groups of stars are
$\sim0.05\Msun$ wide, but the bluest clump spans only $0.01\Msun$ in
mass (but nearly $10,000 {\rm K}$ in $T_{\rm eff}$).  Whatever
physical mechanism is responsible for the presence of the blue tail
must also be capable of producing such astonishingly narrow gaps
between discrete clumps.

The uppermost gap on the blue HB is close to the mass where the HB
evolutionary tracks change from predominantly redward to blueward, as
a result of the decreasing contribution of the hydrogen-burning shell.
This change may account for part of that gap, although it seems likely
that a gap in the mass distribution exists as well.

\section{Discussion}

Why do narrow gaps appear on the horizontal branch of NGC 2808?
Canonical stellar models cannot account for the gaps without appeal to
the mass distribution (Whitney \etal\ 1997, and Fig.\ 3), with the
possible exception of the middle gap.  We must therefore consider the
factors that determine a star's position on the ZAHB.

It is not clear whether the red--blue bimodality necessarily has the
same cause as the gaps within the blue HB.  NGC 1851 shows the
red--blue gap, but not the blue tail.  On the other hand, there is
significant variation in the morphologies of blue HBs.  The HBs of NGC
6752 and M13 (Paltrinieri \etal\ 1997) and M15 (Buonanno \etal\ 1985)
show gaps that are arguably similar to the blue HB of NGC 2808.  Two
gaps have also been found in the blue HB of the field halo stars by
Newell (1973).  Thus, the bizarre appearance of the NGC 2808 HB may be
the result of a combination of effects, some seen elsewhere in lesser
degree, and some unique to this cluster.

Variations in the {\it overall} HB morphology between clusters have
received much attention.  In addition to the ``first parameter''
(metal abundance), a number of ``second-parameter'' candidates have
been proposed.  Much recent debate has centered on whether the
predominant second parameter is age (Lee, Demarque, \& Zinn 1994;
Stetson, VandenBerg, \& Bolte 1996).  Other candidates include He and
CNO abundance (which alter the shell-burning rate), rotation and
number density of stars (which, indirectly, affect the amount of mass
lost by a star), or a combination of candidates.

Most of these candidates do not offer a satisfactory explanation for
the gaps we see here, or for red--blue bimodality in general.  If NGC
2808 had a spread in [Fe/H] or age, we would not expect to see as
narrow an RGB and turnoff, respectively, as are seen in the optical
CMD (Buonanno \etal\ 1984).  The suggestion of van den Bergh (1996)
that cluster mergers are responsible for bimodality appears unlikely
for this same reason.  A peculiar distribution of abundances of other
elements could help to account for the HB structure seen here.
However, CNO abundance variations have not been very successful in
explaining the HB morphologies of other clusters (Dickens \etal\
1991), and the problem of why the distribution was abnormal in NGC
2808 would remain.

Various authors (Renzini 1977, Buonanno \etal\ 1985) have suggested
that stellar rotation might account for bimodality.  Rood \etal\
(1993) proposed that if rotation causes mass loss through some MHD
effect, then there could be a critical rotational speed beyond which
mass loss is enhanced.  Another possibility is suggested by the
hypothesis of Sweigart (1997) of He mixing on the RGB.  Mixing would
lead to bluer HB colors, because of the higher RGB tip luminosity and
thus greater mass loss, and also because of the higher He abundance in
the envelopes of the HB stars.  If mixing took place only above a
critical rotational speed, it might lead to a peculiar HB morphology.
However, rapid rotation would lead to a larger core for the bluer
stars, destroying the agreement of the HB locus and the
constant-core-mass model prediction (Fig.\ 3).

The local environment may play a role.  Clusters with denser cores
often have bluer HBs (Fusi Pecci \etal\ 1993, and references therein),
and tidal stripping of red-giant envelopes during stellar encounters
has been suggested as the cause.  The revised parameters given in
Fig.\ 3 lead to a lower central density for NGC 2808 than previously
thought ($\log \rho_0 = 4.45,$ rather than 4.63 [Djorgovski 1993], in
$L_{\odot V}/{\rm pc}^3$).  This density is still high, and the
presence of the blue tail is consistent with what is seen in other
dense clusters (Buonanno \etal\ 1997).  We estimate that $\sim100$
close encounters have occurred in the cluster in the last Gyr,
although not all of these involved red giants.  If tidal interactions
lead to mass loss, then some stars would, by chance, have multiple
encounters, which would explain the multiple HB populations in a
natural way.

One potential problem with this scenario is that the blue and red HB
stars have identical radial distributions.  Presumably, the
progenitors of the blue HB stars suffered their encounters within the
cluster core.  A Fokker--Planck simulation of NGC 2808 shows that the
orbits of stars initially within $r_c$ do not evolve much in $10^8$ yr
(the HB lifetime).  The lack of radial gradients would thus require
the encounters to occur near the base of the RGB.  (The longer time
spent at that stage would compensate for the stars' smaller
cross-section.)  It is also possible that encounters involving
binaries would expel stripped stars to larger radii.

Also, for narrow gaps to appear on the HB, the encounters must be
``tuned'' to cause particular amounts of mass loss on each
encounter---amounts that are nearly the same from star to star.  It
seems implausible that the sequence of tidal events would be so
reproducible.  Another objection is that the relative numbers of stars
in the groups are not well fit by the predictions of this scenario, in
its simplest form.  If we assume that each star has an equal encounter
probability per unit time (a poor approximation, given that the
density varies widely with $r$), then the counts should follow a
Poisson distribution.  The best-fitting distribution predicts 352,
289, 119, and 40 stars in the four groups, from red to blue.  The last
two predictions disagree strongly with the data:\ the $\chi^2$
distribution gives a probability less than $10^{-6}$ of such a severe
disagreement.  On the other hand, a model that took the varying
density into account might fit more closely.

Finally, stellar encounters could not be the cause of the gaps in the
field-star distribution, although the latter might arise from an
entirely different phenomenon.  The encounter frequency is also much
lower in NGC 6229 and in M13.

The presence of a binary companion could be important.  A somewhat
higher frequency of hot field subdwarfs are found in binaries (Allard
\etal\ 1994), as are a number of the blue stars in NGC 6791 (Green
\etal\ 1997).  The bluest clump of HB stars in $\F218W - B$ has a
slight redward extension in $B-V$, possibly indicating that some of
these stars are blended with a main-sequence companion.  However, the
binary-star hypothesis would also lead us to expect radial population
gradients, which are not observed.

Since the physics of stellar mass loss is so poorly understood, we
must be somewhat speculative in considering further options.  One
plausible hypothesis to explain multimodality is that multiple
mass-loss processes act in red-giant stars.  There is little physical
justification available for any specific mass distribution, and no
survey has addressed this problem in a systematic way.  Some red HBs
could even be multimodal, but this effect could not be observed
because of the small variation of color with envelope mass for red HB
stars.

Mass-loss processes typically considered for solitary cool stars
include molecular-line- and grain-driven winds, by which the material
in the lower atmosphere must couple to the radiation field.  Such
processes might depend on the detailed line formation; if so, the
locations of HB gaps may be consistent between clusters.  Also,
spectroscopy might reveal differences between stars on either side of
a gap (see Peterson, Rood, \& Crocker 1995, Moehler \etal\ 1995).

In summary, we can provide no satisfactory explanation for the
multimodality of the horizontal branch of NGC 2808.  These
observations serve to remind us that stellar mass loss is a poorly
understood phenomenon that awaits much further work.  Our hope is that
this work will eventually tell us why NGC 2808, out of all
well-observed globular clusters, is blessed with a combination of
factors that gives its HB such a spectacular appearance.

\acknowledgements

This work was supported by NASA grant GO--6095 from STScI, as well as
by NASA grants NAG5-700 and NAGW-4106 (BD), NAG5-2756 (ESP), by the
Bressler Foundation (SGD), and by the Agenzia Spaziale Italiana and by
the Ministero dell'Universit\`a e della Ricerca Scientifica e
Tecnologica (GP and AR).

\clearpage

%
%

\clearpage

%
%

\def\capone{$(V,B-V)$ CMD of 27,286 stars in NGC 2808.  Note the
extended blue HB, and the prominent blue straggler sequence.}

\def\captwo{{\it Lower panel:} $(B,\F218W-B)$ CMD of 437 blue stars in
NGC 2808.  $B$ magnitudes are in the Johnson system, and F218W
magnitudes are in the STMAG system.  The reddest portion of the HB
does not appear on this diagram.  The larger symbols are the HB stars
that are histogrammed in the upper panel; smaller symbols are objects
not on the HB. {\it Upper panel:} Histogram of colors of blue HB
stars.  Note the two gaps.}

\def\capthree{A fit of ZAHB models and isochrones to the CMDs.  The upper
panel shows the Johnson $(B,\> V)$ fit, with observed colors shifted
as described in the text.  The lower panel shows the $(\F218W,\> B)$
fit.  The line along the HB connects ZAHB models of Dorman, Rood, \&
O'Connell (1993), where each actual model is shown as a circle, and
effective temperatures of these models are shown along the top of the
lower panel.  The narrow dark lines rising from the ZAHB are
evolutionary tracks, from the same paper, for HB stars of masses
0.4848, 0.495, 0.510, 0.530, and 0.540 $\Msun$ (from blue to red).
The other heavy lines are a zero-age main sequence for stars with
masses $< 1.5 \Msun$, and an isochrone with [Fe/H]$ = -1.48$ and $t =
14$ Gyr (the age was not fitted to the data).}

%
%




%
%

\clearpage

\begin{figure}
\figurenum{1}

\plotone{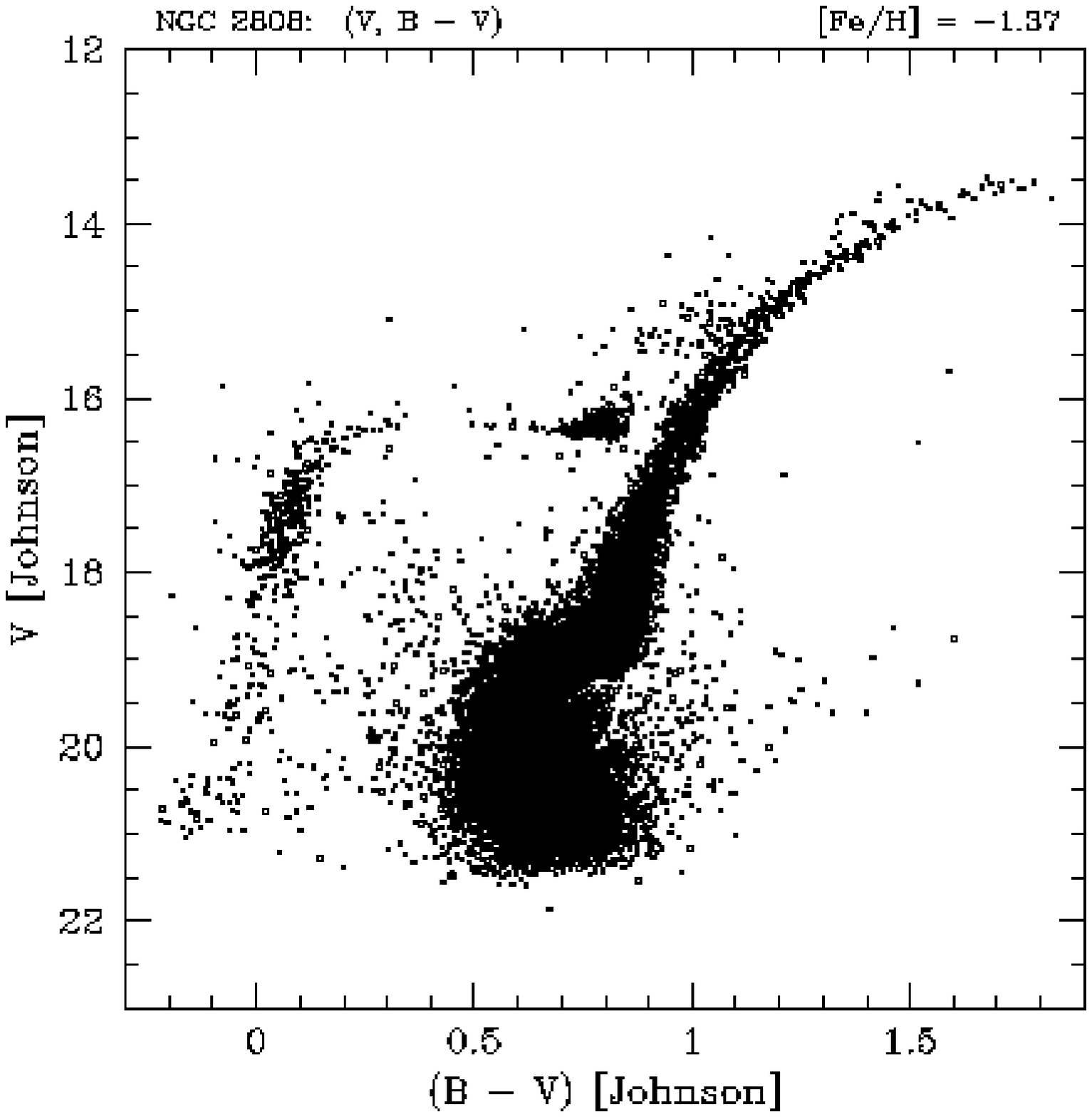}
\caption{\capone}
\label{bv_cmd}
\end{figure}

\clearpage

\begin{figure}
\figurenum{2}

\plotone{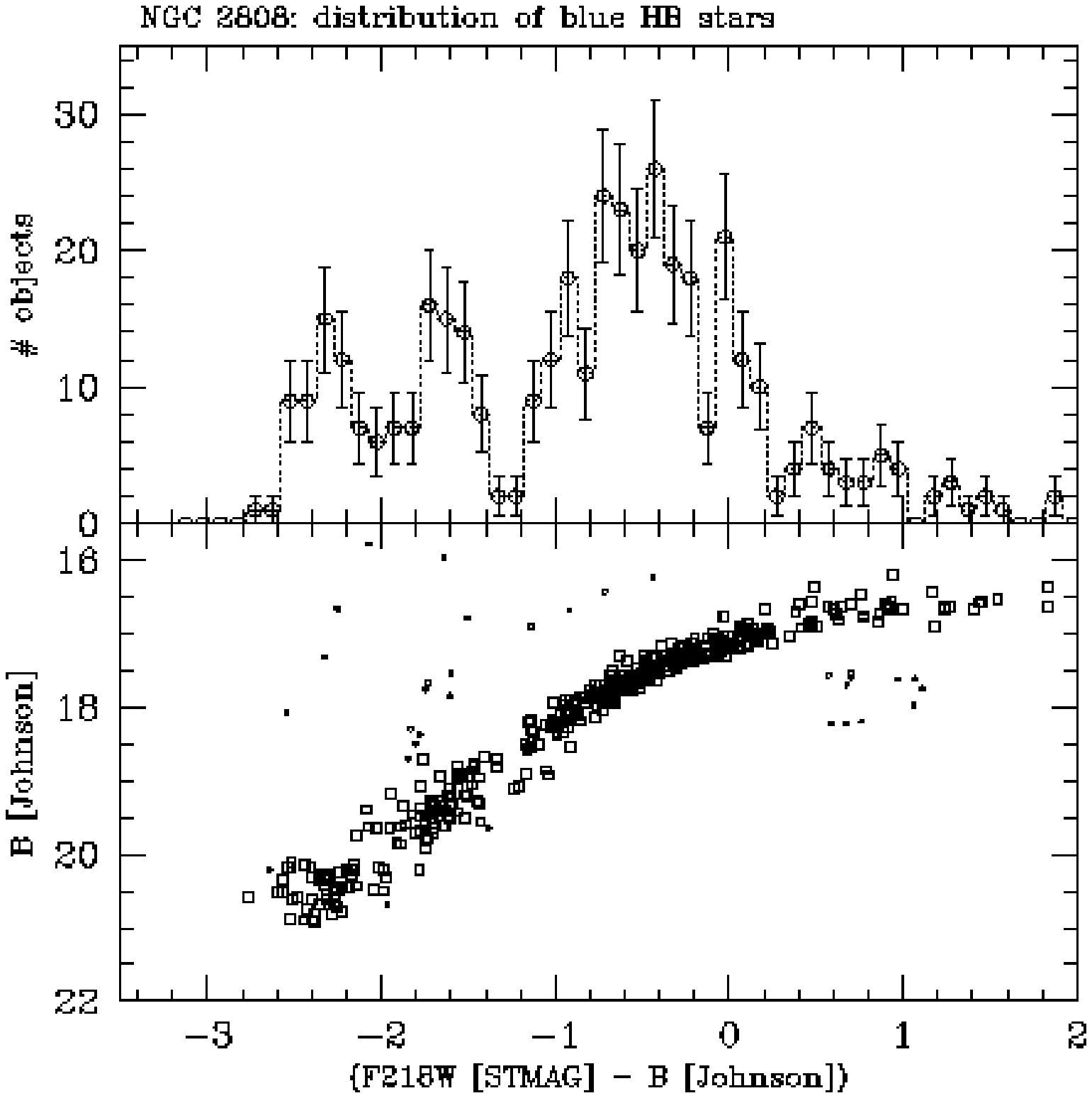}

\caption{\captwo}
\label{bhb_dist}
\end{figure}

\clearpage

\begin{figure}
\figurenum{3}

\plotone{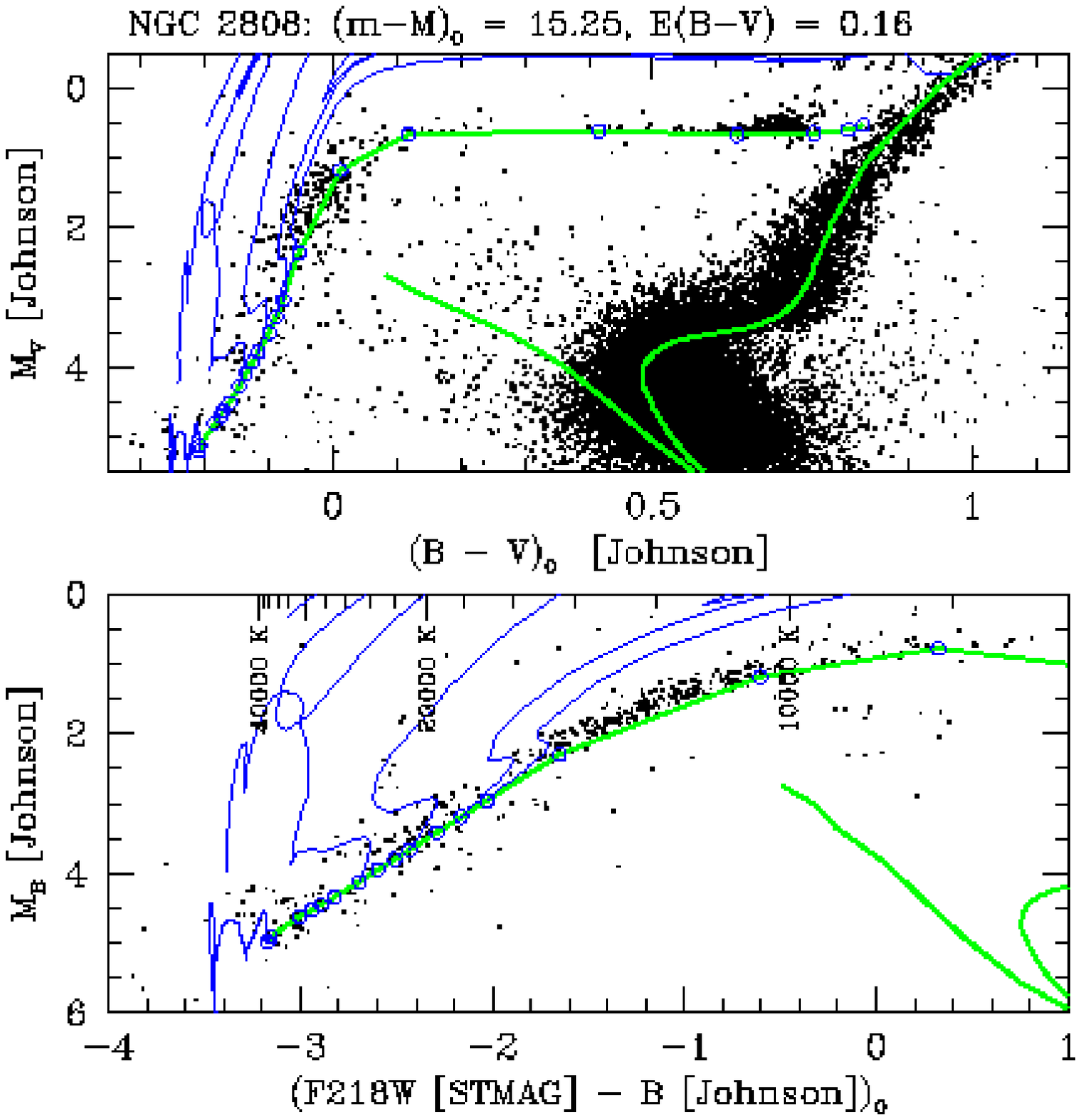}

\caption{\capthree}
\label{fit}
\end{figure}


\begin{thebibliography}{DUMMY}

\bibitem[Allard \etal\ 1994]{al94}
Allard, F., Wesemael, F., Fontaine, G., Bergeron, P., \& Lamontagne,
R.\ 1994, \aj, 107, 1565

\bibitem[Borissova \etal\ 1997]{bor97}
Borissova, J., Catelan, M., Spassova, N., \& Sweigart, A.\ V.\ 1997,
to appear in \aj, February 1997

\bibitem[Buonanno \etal\ 1984]{buo84}
Buonanno, R., Corsi, C.\ E., Fusi Pecci, F., \& Harris, W.\ E.\ 1984,
\aj, 89, 365

\bibitem[Buonanno \etal\ 1985]{buo85}
Buonanno, R., Corsi, C.\ E., \& Fusi Pecci, F.\ 1985, 
\aap, 145, 97

\bibitem[Buonanno \etal\ 1997]{buo97} 
Buonanno, R., Corsi, C.\ E., Bellazzini, M., Ferraro, F.\ R., \& Fusi
Pecci, F.\ 1997, to appear in AJ, February 1997

\bibitem[Byun \& Lee 1991]{byun}
Byun, Y.-I., \& Lee, Y.-W.\ 1993, in The Formation and Evolution of
Star Clusters, ASP Conf.\ Series Vol.\ 13, ed.\ K.\ Janes (San
Francisco:\ ASP), p.\ 243

\bibitem[Cardelli, Clayton, \& Mathis 1989]{ccm89} 
Cardelli, J., Clayton, G.\ C., \& Mathis, J.\ S.\ 1989, \apj, 345, 245

\bibitem[Castellani \& Renzini 1968]{cr68}
Castellani, V., \& Renzini, A.\ 1968, Astrophys.\ Space Sci., 2,
310

\bibitem[Crocker, Rood, \& O'Connell 1988]{cro88}
Crocker, D.\ A., Rood, R.\ T., \& O'Connell, R.\ W.\ 1988, \apj, 332, 236

\bibitem[Dickens, Croke, Cannon, \& Bell 1991]{dcb91}
Dickens, R.\ J., Croke, B.\ F.\ W., Cannon, R. D., \& Bell, R.\ A.\
1991, Nature, 351, 212

\bibitem[Djorgovski 1993]{djtable}
Djorgovski, S.\ 1993, in Structure and Dynamics of Globular Clusters,
ASP Conf.\ Series, Vol.\ 50, eds.\ S.\ Djorgovski \& G.\ Meylan (San
Francisco:\ ASP), p.\ 373

\bibitem[Dorman, Rood, \& O'Connell 1993]{dro93}
Dorman, B., Rood, R.\ T., \& O'Connell, R.\ W.\ 1993, \apj, 419, 596

\bibitem[Ferraro \etal\ 1990]{fer90}
Ferraro, F.\ R., Clementini, G., Fusi Pecci, F., Buonanno, R., \&
Alcaino, G.\ 1990, Astron.\ Astrophys.\ Suppl., 84, 59

\bibitem[Ferraro \etal\ 1997]{fer97}
Ferraro, F., Paltrinieri, B., Fusi Pecci, F., Cacciari, C., Dorman,
B., \& Rood, R. T.\ 1997, submitted to ApJL

\bibitem[Fusi Pecci \etal\ 1993]{fusipecci}
Fusi Pecci, F., Ferraro, F.\ R., Bellazzini, M., Djorgovski, S.\ G.,
Piotto, G., \& Buonanno, R.\ 1993, \aj, 105, 1145

\bibitem[Green \etal\ 1997]{gre97}
Green, E.\ M., Liebert, J., Peterson, R.\ C., \& Saffer, R.\ A.\ 1997,
in The Third Conference on Faint Blue Stars, eds.\ A.\ G.\ D.\ Philip,
J.\ Liebert, \& R.\ A.\ Saffer (Schenectady:\ L.\ Davis Press), in
press

\bibitem[Harris 1974]{harris}
Harris, W.\ E.\ 1974, \apjl, 192, L161

\bibitem[Hill \etal\ 1996]{hill}
Hill, R.\ S.\ \etal\ 1996, \aj, 112, 601

\bibitem[Holtzman \etal\ 1995]{holtzman}
Holtzman, J.\ A., Burrows, C.\ J., Casertano, S., Hester, J.\ J.,
Trauger, J.\ T., Watson, A.\ M., \& Worthey, G.\ 1995, \pasp, 107, 1065

\bibitem[Iben \& Rood 1970]{ir70}
Iben, I., \& Rood, R.\ T.\ 1970, \apj, 161, 587 

\bibitem[Kaluzny \& Udalski 1992]{ku92}
Kaluzny, J., \& Udalski, A.\ 1992, Acta Astronomica, 42, 29

\bibitem[Kurucz 1992]{kur}
Kurucz, R.\ L.\ 1992, in Precision Photometry:\ Astrophysics of the
Galaxy, eds.\ A.\ G.\ D.\ Philip, A.\ R.\ Upgren, \& K.\ A.\ Janes
(Schenectady: L.\ Davis Press), p.\ 27

\bibitem[Lee, Demarque, \& Zinn 1994]{ldz94}
Lee, Y.-W., Demarque, P., \& Zinn, R.\ 1994, \apj, 423, 248

\bibitem[Liebert, Saffer, \& Green 1994]{lsg94}
Liebert, J., Saffer, R.\ A., \& Green, E.\ M.\ 1994, \aj, 107, 1428

\bibitem[Moehler \etal\ 1995]{moehler}
Moehler, S., Heber, U., \& Deboer, K.\ S.\ 1995, \aap, 294, 65

\bibitem[Newell 1973]{newell}
Newell, E.\ B.\ 1973, \apjs, 26, 37

\bibitem[Peterson 1993]{table}
Peterson, C.\ 1993, in Structure and Dynamics of Globular Clusters,
ASP Conf.\ Series, Vol.\ 50, eds.\ S.\ Djorgovski \& G.\ Meylan (San
Francisco:\ ASP), p.\ 337

\bibitem[Peterson, Rood, \& Crocker (1995)]{psc}
Peterson, R.\ C., Rood, R.\ T., \& Crocker, D.\ A., \apj, 453, 214

\bibitem[Renzini 1977]{ren77}
Renzini, A.\ 1977, in Advanced Stages in Stellar Evolution, eds.\ P.\
Bouvier \& A.\ Maeder (Sauverny:\ Geneva Obs.), p.\ 149

\bibitem[Rich \etal\ 1997]{us}
Rich, R.\ M., \etal\ 1997, ``Extended Blue Horizontal Branches in 
Metal-Rich Globular Clusters,'' submitted to ApJL

\bibitem[Rood \etal\ 1993]{rood93}
Rood, R.\ T., Crocker, D.\ A., Fusi Pecci, F., Ferraro, F.\ R.,
Clementini, G., \& Buonanno, R.\ 1993, in The Globular Cluster--Galaxy
Connection, ASP Conf.\ Series, Vol.\ 48, eds.\ G.\ H.\ Smith \& J.\
P.\ Brodie (San Francisco:\ ASP), p.\ 218

\bibitem[Stetson 1981]{pbs81}
Stetson, P.\ 1981, \aj, 86, 687

\bibitem[Stetson 1987]{pbs87}
Stetson, P.\ 1987, \pasp, 99, 191

\bibitem[Stetson, VandenBerg, \& Bolte 1996]{svb96}
Stetson, P., VandenBerg, D.\ A., \& Bolte, M.\ 1996, \pasp, 108, 560

\bibitem[Sweigart 1997]{swe9}
Sweigart, A.\ V.\ 1997, \apjl, 474, L23

\bibitem[van den Bergh (1996)]{vdb96}
Van den Bergh, S.\ 1996, \apjl, 471, L31

\bibitem[Whitmore, Heyer, \& Baggett 1996]{wfpcISR}
Whitmore, B., Heyer, I., \& Baggett, S.\ 1996, ``Effects of
Contamination on WFPC2 Photometry,'' WFPC2 Instrument Science Report
96-4 (Baltimore:\ STScI)

\bibitem[Whitney \etal\ 1997]{whit97}
Whitney, J. H., \etal\ 1997, in preparation

\bibitem[Yi, Lee, \& Demarque 1993]{yld93}
Yi., S., Lee, Y.-W., \& Demarque, P.\ 1993, \apj, 411, L25

\end{thebibliography}
\end{document}